\documentclass[a4paper,epj]{svjour}

\usepackage{graphicx}
\usepackage{verbatim}
\usepackage{fancyhdr}
\usepackage{cite,mcite}
\usepackage[small]{caption}
\usepackage{amssymb}
\def\makeheadbox{{
 }}

\def\mytitle{My title} 
\def\myauthors{My name}  
\def\mytype{My type of session}
\def\mysession{My session}

\def\mytitle{Supersymmetric LHC phenomenology without a light Higgs boson}
\def\myauthors{Roberto Franceschini}
\def\mytype{Contributed Talk}    
\def\mysession{Colliders - SUSY Phenomenology}

\begin{document}
\title{Supersymmetric LHC phenomenology without a light Higgs boson}
\author{Roberto Franceschini
\thanks{\emph{Email:} r.franceschini@sns.it} 
}                     
%
%
\institute{Scuola Normale Superiore and INFN, Piazza dei Cavalieri 7 - I-56126 Pisa, ITALIA}
%
\date{}
\abstract{
After a brief discussion of the mass of the Higgs in supersymmetry, I introduce $\lambda$SUSY, a model with an extra chiral singlet superfield in addition to the MSSM field content. The key features of the model are: the superpotential $W=\lambda S H_d H_u$ with a large coupling $\lambda$ and the resulting lightest Higgs with mass above 200GeV. The main part of my contribution will be about how $\lambda$SUSY manifests itself at the LHC. Discoveries of gluino, squarks and in particular of the three lightest neutral Higgs bosons are discussed.%
} 
\maketitle
\section{Introduction\label{sec:Introduction}}
%
\quad Soon the LHC will exploit its potential for a early discovery
of a Higgs boson and, in case it is heavier than 140 GeV, the MSSM
will be ruled out. The same conclusion applies to the majority of supersymmetric
models, which, under the assumption of perturbative gauge coupling
unification, cannot have a Higgs heavier than 200 GeV\cite{quiros}.
This generic lightness of the Higgs can receive some support from
the EWPT, which find $m_{h}=76_{-24}^{+33}$ GeV \cite{LEPEWWG} at 65\%
C.L in the SM. However, this result could be misleading for beyond
standard model (BSM). Indeed the SM result extends to BSM only with
the assumption that the new physics, while cutting off top (and
gauge boson) loops, does not itself contribute significantly to the EWPT parameters
$S$ and $T$. Scenarios of BSM not respecting this assumption have been
realized in simple explicit models \cite{improved}, finding an interesting
and alternative LHC phenomenology. In this contribution I will
deal with the phenomenology of one of these alternative scenarios,
$\lambda$SUSY, which in this discussion seems particularly motivated.
In fact it's supersymmetric and has a lightest Higgs naturally above 200 GeV \cite{lsusy}.

The field content of $\lambda$SUSY is that of the MSSM plus a chiral singlet
superfield $S$. The key feature of the model is the presence of the
superpotential interaction\[
W=\lambda S H_{d} H_{u}\] with a large coupling $\lambda$. The maximal value of $\lambda$
is limited by the assumption that it stays perturbative up to about
$10$ TeV, so that the incalculable contribution to the EWPT from
the cutoff can be neglected \footnote{Taking $\lambda=2$ at
the Fermi scale, the Landau pole is at about 50 TeV, which can be interpreted as the compositeness
scale of (some of) the Higgs bosons \cite{fat},\cite{UV}. Ref. \cite{UV} also provides a UV completion compatible with gauge coupling unification.}.%
\section{The model}%
The full $\lambda$SUSY superpotential and the resulting scalar potential are: 
\begin{eqnarray*}
&W=\mu(S)H_{1} H_{2}+f(S),\quad \lambda=\mu^{\prime}(S),\\
&V=\sum_{i}\mu_{i}^{2}(S)|H_{i}|^{2} -(\mu_{3}^{2}(S)H_{1}H_{2}+\mathrm{h.c.})+\\ 
& +\lambda^{2}|H_{1}H_{2}|^{2}+V(S)+...~,
\end{eqnarray*}
\begin{figure}[ptb]
\begin{center}
\includegraphics[clip,height=0.33\textwidth,angle=0]{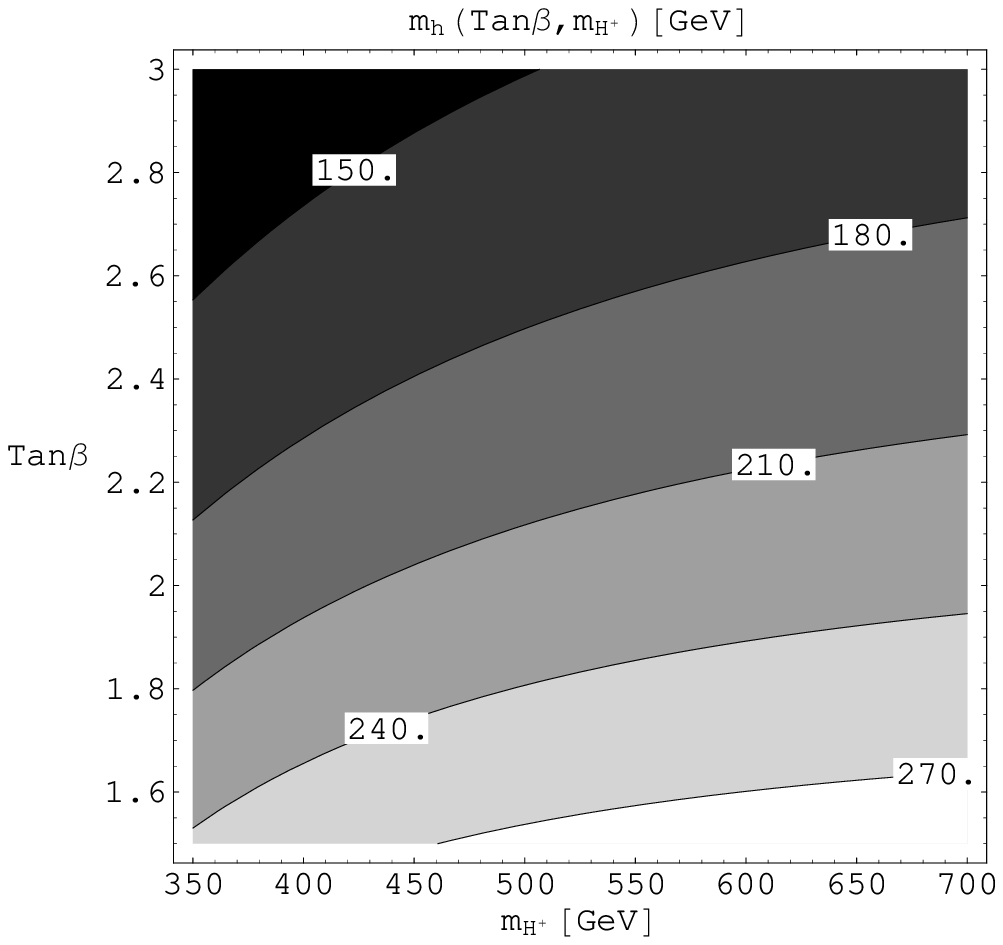} 
\includegraphics[clip,height=0.25\textwidth,angle=0]{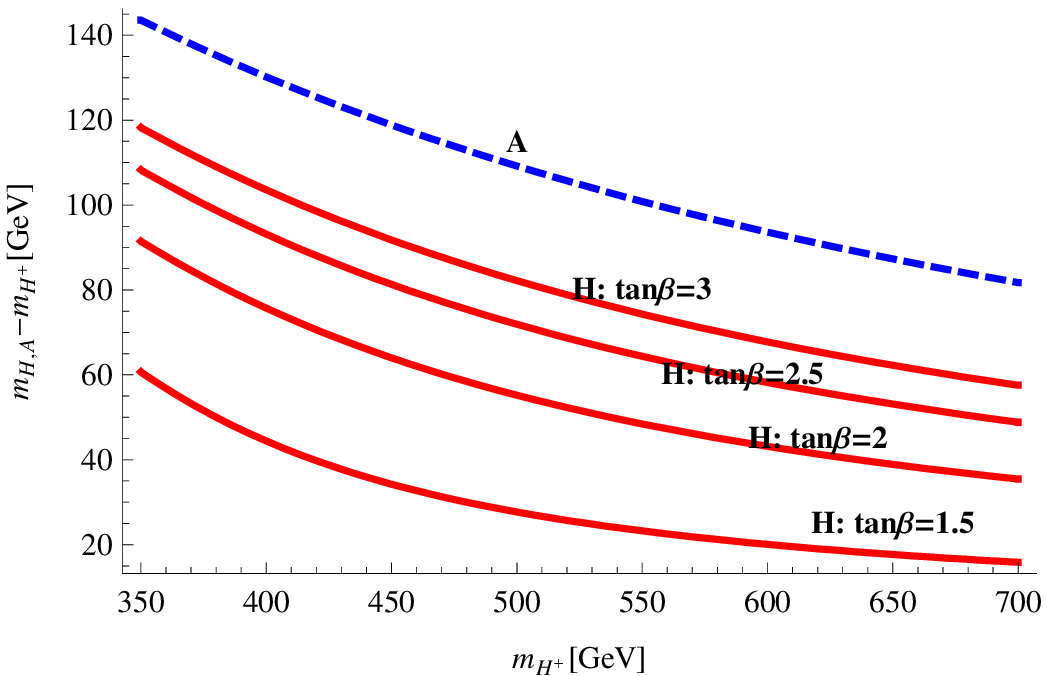} 
\end{center}
\caption{$m_{h}$ (up) and the mass differences $m_{H}-m_{H^{\pm}}$ (down,solid
red lines, $\tan\beta=1.5,2,2.5,3$) and $m_{A}-m_{H^{\pm}}$
(down, dashed blue line) as a function of $m_{H^{\pm}}$ and $\tan\beta$ in the preferred
region (\ref{tan}) of the parameter space for$\lambda=2$.}
\label{masse}
\end{figure}
where the dots stand for negligible D-terms. Assuming the scalar $S$
is heavy and not mixed, the mass eigenstates are the same as in the
MSSM: two CP-even bosons, $h$ and $H$, one CP-odd pseudo-scalar, $A$, and one charged Higgs, $H^{+}$
\footnote{Analysis of concrete examples shows that singlet admixture in $h,H,A$ typically stays below $0.2-0.3$.%
}. Using results of \cite{lsusy}, the spectrum can be given in terms of $m_{H^{+}}$,
$\tan\beta$ and $\lambda$. For $\lambda=2$, the EWPT, Naturalness and  $b\rightarrow s\gamma$ determine the preferred parameter space \cite{lsusy,Gambino}:\begin{equation}
1.5\lesssim\tan\beta\lesssim3, \quad 350 {\rm GeV}\lesssim m_{H^{\pm}}\lesssim700 {\rm GeV.}\label{tan}\end{equation}
Masses of neutral scalars in this range of parameters
are given in Fig. \ref{masse}. The key feature of the spectrum is
that the lightest Higgs boson $h$ is in the $200-300$GeV range,
hence much heavier than in MSSM or NMSSM. Another notable feature
is the fixed ordering of the spectrum: $m_{h}<m_{H^{+}}<m_{H}<m_{A}\ $
(see Fig. \ref{masse}).

It should be noticed that this model also has a Higgsino-like DM candidate (see details in \cite{lsusy}). 
\section{Early (puzzling) discoveries}
As in more standard supersymmetric models, also in $\lambda SUSY$
there are strongly-interacting superpartners. The $O({\rm pb})$
cross section makes them interesting candidates for early discovery
of SUSY at the LHC. As explained below, also the Higgs boson $h$
should be discoverable in the early phase.

{\bf Gluino and stop}\quad Naturalness can be used to bound gluino and stop masses. According to \cite{lsusy} they have to satisfy \[
m_{\tilde{t}}\lesssim800{\rm GeV,} \quad m_{\tilde{g}}  \lesssim1.6{\rm TeV,} \]
while the masses of the electroweak gauginos, sleptons and all the
other squarks, do not have significant bounds. Even assuming
the less favorable scenario where only $\tilde{t}$ and $\tilde{g}$
are light enough to be produced, we expect a early discovery of these
superpartners through their usual cascade decays \cite{ATLAS2,CMS2}.
For a rough estimate of
the discovery potential we can use the existing study \cite{bityukov} valid in the case of effective supersymmetry \cite{effective}. We conclude that $10 {\rm fb}^{-1}$ of integrated
luminosity ($\mathcal{L}$) should be enough for a discovery of SUSY in the entire
range of stop and gluino masses suggested by Naturalness.

{\bf The lightest Higgs}\quad The light Higgs mass is about 200-300 GeV and it's coupling to SM
particles are very close to the SM Higgs value. In \cite{CFR} has been
found that the impact of couplings to superpartners is very limited.
Thus, we estimate the discovery potential of this Higgs boson using
standard model studies \cite{ATLAS2},\cite{CMS2}. We expect an early
discovery of this Higgs boson in the {}``gold-plated'' channel $h\rightarrow ZZ\rightarrow l^{+}l^{-}l^{+}l^{-}$,
with $\mathcal{L}=5{\rm fb}^{-1}$.

For what we have said in the Introduction, the discovery of such a
heavy Higgs boson together with the discovery of superpartners could
be puzzling. $\lambda SUSY$ is a possible solution to this puzzle,
therefore we should look closer to the heavier scalars and ask ourselves
if their detection can give an experimental evidence for this model.
To this aim we assume $h$, $\tilde{g}$ and $\tilde{t}$ have been
observed and $m_{h}$ is known, then we turn to study the discovery
reach for H and A.
\section{Investigation of $\lambda$SUSY}
{\bf The heavy CP-even scalar}\quad Form Fig. \ref{masse} we see that the heavy CP-even Higgs boson $H$ has mass in the 500-800 GeV range. Interactions of H are described in \cite{CFR} where the following results were found. The H is a quite narrow resonance, its width ranges from 5 to 40 GeV. Whenever it is kinematically available, there is a dominance of the $H\to hh$ decay mode. This can be ascribed to the large $\lambda$, which enters quadratically in the expression for the $Hhh$ coupling. The stop gets decoupled as is gets heavier and assuming $m_{\tilde{t}_{1}}>400{\rm GeV}$ one can neglect couplings to stop.  
In \cite{CFR,lsusy} interactions with Higgsinos have been studied. They depend on $\mu$ and on the mass of the heavy scalar S. In our concrete study we take a small value for the $H\chi\chi$ coupling, that is we maximize the
branching fraction of $H$ into standard model particles. This can be thought as a favorable condition for $H$ discovery%
\footnote{Inclusion of $H$ decays into Higgsinos lessens the final result at
worse by a factor $0.5$. %
}. 

To assess LHC's potential for H discovery we choose a rather generic
point of the parameter space (\ref{tan}): 
\begin{equation}
\tan\beta=2,\quad m_{H^{+}}=500{\rm GeV,}\label{punto}
\end{equation}
and study possible detection strategies. Relevant particle properties
for the choice of parameters Eq. (\ref{punto}) are given in Table
\ref{tab:Particle-properties}. For plots of these quantities in the
whole parameter space Eq. (\ref{tan}) we refer to \cite{CFR}. %
\begin{table}[!h]
\begin{tabular}{|c|c|c|c|c|}
\hline 
$\sigma_{H}^{GF}$ & $m_{H}$ & $\Gamma_{H}$ & $m_{h}$ & $\Gamma_{h}$\tabularnewline
\hline 
150 fb & 555 GeV & 21 GeV & 250 GeV & 3.8 GeV\tabularnewline
\hline
\hline 
$\sigma_{H}^{VBF}$ & $\frac{g_{Htt}^{2}}{g_{Htt,SM}^{2}}$ & $\frac{g_{HVV}^{2}}{g_{HVV,SM}^{2}}$ & $\mathcal{B}_{(H\rightarrow hh)}$ & $\mathcal{B}_{(H\rightarrow VV)}$\tabularnewline
\hline 
27 fb & 0.058 & 0.060 & 0.76 & 0.2\tabularnewline
\hline
\hline 
$\sigma_{A}^{GF}$ & $m_{A}$ & $\Gamma_{A}$ & $\mathcal{B}_{(A\rightarrow hZ)}$ & $\mathcal{B}_{(A\rightarrow t\bar{t})}$\tabularnewline
\hline 
0.7pb & 615GeV & 11 GeV & 0.2 & 0.76 \tabularnewline
\hline
\end{tabular} 

\caption{\label{tab:Particle-properties}Particle properties at the point Eq.
(\ref{punto}). VBF means vector boson fusion. $\mathcal{B}$ means branching fraction. $g_{,SM}$ are the couplings of a same mass SM Higgs boson.}

\end{table}

From Table \ref{tab:Particle-properties} we see that $H$ is mainly
produced via gluon fusion (GF), thus, in the following we will consider
only this channel. 

Once produced, most of the $H$s will decay into $hh$ and then into
$4V$, resulting in $\sigma_{tot}(gg\rightarrow H\rightarrow4V)=110 {\rm fb}$ ($V$ means both $Z$ and $W$).
To have a sizeble final state cross section we cannot demand more
than one leptonic decay of these weak bosons. Our choice for a quantitative
study is therefore:
\begin{equation}
gg\rightarrow H\rightarrow hh\rightarrow2Z2V\rightarrow l^{+}l^{-}6J,\label{our-channel}
\end{equation}
Signal is defined with $J=\{u,d,c,s,b,g\}$ and has been produced with \textsc{madgraph} \cite{Maltoni:2002qb}, which yields \[
\sigma\times BR=2.67{\rm fb.}\]

For the $H$ mass values we are interested in, the relevant background (BG)
sources of $l^{+}l^{-}6J$ events are the $Z6J$ and $t\bar{t}Z$ %
\footnote{ All the details about the identification of relevant BG sources
and their simulation can be found in \cite{CFR}.%
}. The latter has been simulated with \textsc{madgraph} while for $Z6J$
we used specific\textsc{ alpgen}\cite{Mangano:2002ea} codes for $Z6j$ and $ZQ\bar{Q}4j$ ($Q=b,c$).

{}Restricting event invariant mass in a $O(100 {\rm GeV})$ interval
around $m_{H}$, the total BG cross section is a factor $2000$
bigger than that of the signal. To increase the S/B ratio we exploit the presence of intermediate resonances ($h,W,Z)$ in the signal. Thus we enforce \textit{reconstruction cuts} on both signal and BG, i.e. we require that the intermediate state resonances be reconstructed
by final state jets and leptons. This is the main tool we use the in our analysis. 

All samples are analyzed with \textsc{root} \cite{ROOT}. We don't do neither showering nor jet reconstruction simulation, in fact our analysis is completely partonic. We also ignore flavor tagging and trigger issues, but our inclusive definition of jet, J, and final selection cuts Eq. (\ref{eq:cuts}),
respectively, make these simplifications fully justified. However,
in order to make the analysis more realistic, we do introduce a smearing
of energies of individual jets. The smearing coefficient is generated using the expression\footnote{This is one of the values discussed in Table 9-1 of \cite{ATLAS2}.} $
\sigma/E=0.5/\sqrt{E/{\rm GeV}}+0.03$.
After smearing, we impose the kinematical cuts:\begin{eqnarray}
\Delta R_{JJ}>0.7, & p_{T}^{J}>20{\rm GeV,} & \eta_{J}<2.5,\label{eq:cuts}\\
\Delta R_{lJ}>0.4, & p_{T}^{l}>10{\rm GeV,} & \eta_{l}<2.5,\nonumber \\
 & |m_{ll}-m_{Z}|<10{\rm GeV,}\nonumber \end{eqnarray}
where $m_{ll}$ denotes leptons pair invariant mass.

The signal events passing these cuts correspond to $0.42(1)$
fb cross section while BGs cross section is still orders
of magnitudes larger. Finally, we impose the \textit{reconstruction
cuts}, proceeding as follows.%

\textbf{R1.} For each event we try to group the 6 final jets into
3 pairs so that the jets in each pair \textit{reconstruct} a W or
a Z. By this we mean that the invariant mass $m_{inv}$ of each pair
has to satisfy the requirement:\begin{eqnarray}
|m_{inv}-M_{V}|\leq\delta_{V}, & \delta_{V}=8{\rm GeV}, & V\in\{W,Z\}\label{deltaV}\end{eqnarray}

\textbf{R2.} If a grouping into jet pairs reconstructing a W or a
Z each is found, we proceed to impose a further condition that two
$h$'s be reconstructed by four jets from two of these three pairs,
say pair 1 and 2, and by two jets of pair 3 and the two leptons. In
this case the precise reconstruction cut that we used is\begin{eqnarray}
&|m_{pair_{1}+pair_{2}}-m_{h}|\leq\delta_{h},\quad\delta_{h}=18{\rm GeV,}\label{deltah}\\ 
&|m_{pair_{3}+l^{+}l^{-}}-m_{h}|\leq \delta_{h}/\sqrt{2},\nonumber\end{eqnarray}
 where $m_{pair_{1}+pair_{2}}$ and $m_{pair_{3}+l^{+}l^{-}}$ are
the invariant masses of the $4J$ and $2Jl^{+}l^{-}$ final states 
\footnote{The used value of $\delta_{h}$($\delta_{V}$) is motivated by the natural width
of $h$($V$) and jet energy resolution.
}. We also check that the gauge boson reconstructed by the jets of pair 
3 is a Z, while the two gauge bosons reconstructed by the jets of
pairs 1 and 2 are of the same type (both W or both Z).  If no grouping
of 6 jets into 3 pairs satisfying both R1 and R2 can be found (we
go over all combinations), the event is rejected. %
%
%
%
%
\begin{figure}[ptb]
\includegraphics[height=0.28\textwidth,angle=0]{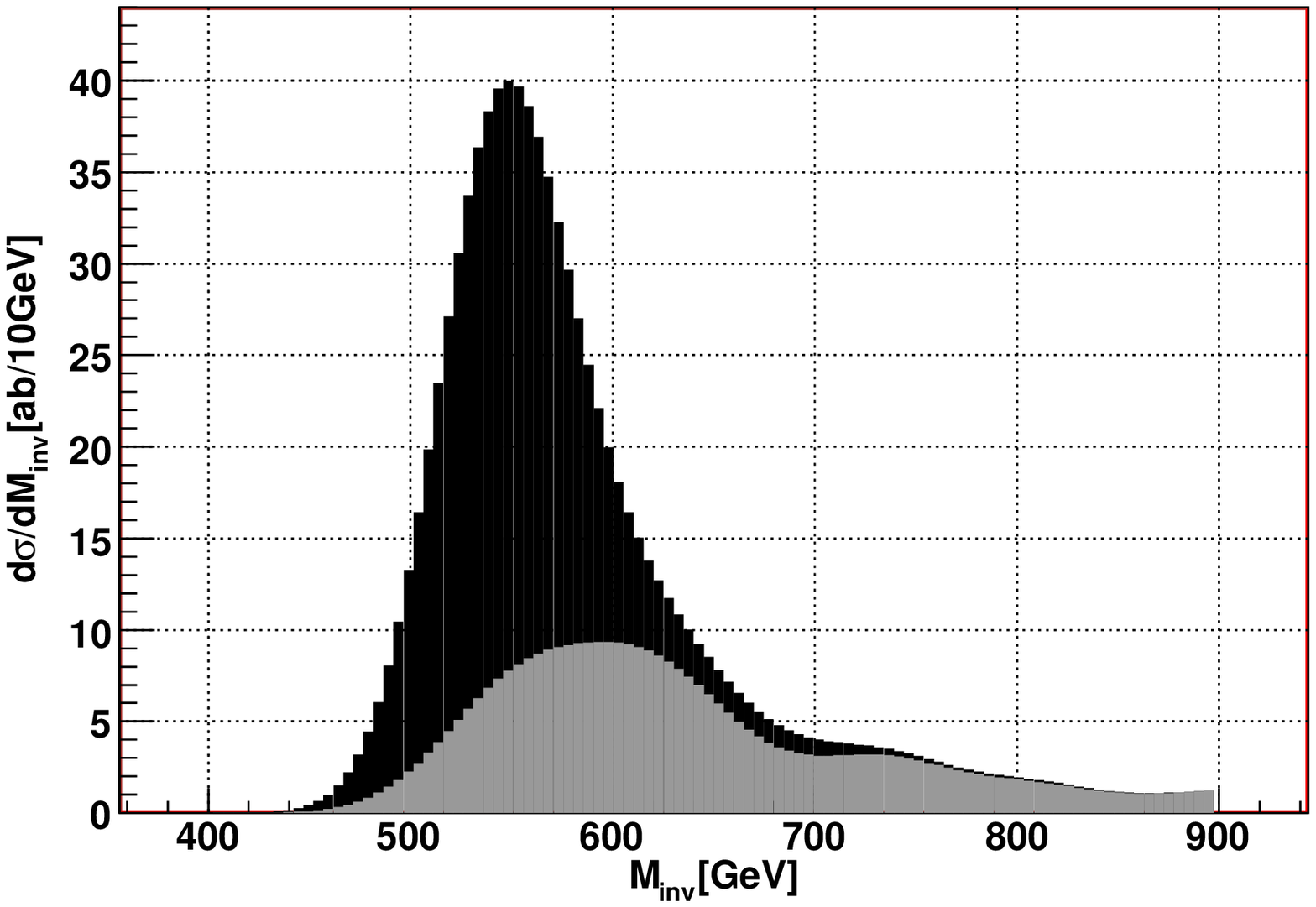} 
\includegraphics[height=0.28\textwidth,angle=0]{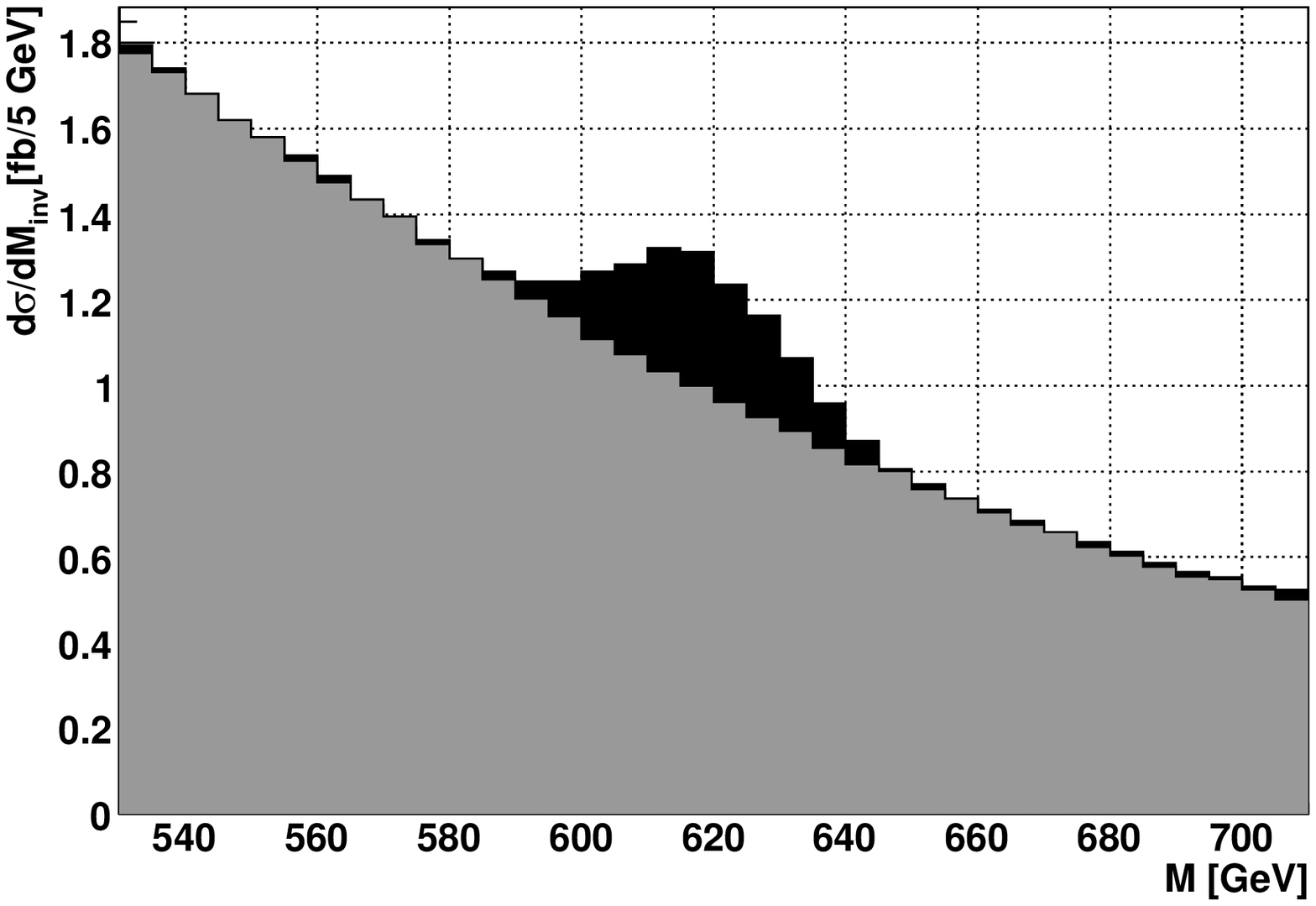} 
\caption{Differential cross section against event invariant mass. Black is $\lambda$SUSY at the benchmark point (\ref{punto}), while grey is the SM background. The upper plots refer to the process (\ref{our-channel}), while the lower plot refers to the process (\ref{Achannel}).}
\label{fig:Signal+BGs-and-BG}
\end{figure}

We ran the reconstruction analysis on the signal sample and on each
of the relevant BG samples. Resulting total cross sections
\textit{after the reconstruction cuts} are 0.286(9)fb and 0.40(1)fb, respectively. Distribution of the signal+BG and of BG-only cross section
versus the total invariant mass of the event is shown in the upper panel of Figure \ref{fig:Signal+BGs-and-BG}.
From this figure it is apparent that the rejection efficiency
of our procedure is high enough to unveil the signal. In particular, we see that
signal and BG peak in the same invariant mass range. The
discovery of $H$ will thus come from an overall excess
of events compared to the SM prediction, as well as from the enhanced
prominence of the SM peak. For $\mathcal{L}=100{\rm fb}^{-1},$
the expected number of events passing all the cuts is $20$ in the
SM, and $49$ in $\lambda$SUSY at the benchmark point (\ref{punto}),
giving $3.4\sigma$ if one uses the significance estimator given in Eq. (A.3) of \cite{CMS2}. Of course, once this global excess
is found, it is worth to scan the invariant mass range to find where
the excess is localized. For instance, for $510$ GeV$<M_{inv}<590$
GeV we have 3 events in the SM, and 23 events in $\lambda$SUSY, $7.2\sigma$
away from the SM. When going beyond benchmark-point analysis (something
out our aim), such localized excess can be used to determine $m_{H}$.%

{\bf The CP-odd pseudo-scalar}
\quad $A$ has mass in the 500-800 GeV
range, as the heavy scalar $H$, but it is always heavier than $H$ (see
Fig. \ref{masse}). The expression for its couplings to the SM fermions
are the same as in the MSSM and can be found in \cite{Djouadi:2005gj}. 
{}By CP-invariance $AVV$ couplings vanish, therefore the only relevant
production mechanism is gluon fusion via the top loop. 

Its total width ranges between 5 and 30 GeV and is dominated
by $A\rightarrow t\bar{t}$ and $A\rightarrow hZ$ decays. Although
the branching ratio of $A\rightarrow t\bar{t}$ is almost always dominant, it is not usable to discover A \cite{top}. Therefore, we focus on $A\rightarrow hZ$, whose BR is smaller, but still significant.

Most of the produced $h$'s will decay into vectors, yielding $\sigma_{tot}(gg\rightarrow A\rightarrow ZVV)\sim100 {\rm fb}$
over all the parameter space. Such a cross section will give too small
event rate if more than one $V$ is allowed to decay leptonically.
Therefore we concentrate on the signature\begin{equation}
gg\rightarrow A\rightarrow hZ\rightarrow VVZ\rightarrow4Jl^{+}l^{-}{\rm \quad(signal)}.\label{Achannel}\end{equation}
We fix the point of parameter space Eq. (\ref{punto}) and compute total cross section for signal (\ref{Achannel})  \[
\sigma\times BR({\rm signal})=6.9{\rm fb.}\]
Signal events has been produced with \textsc{madgraph}.

{}The relevant BG sources of $l^{+}l^{-}4J$ events are the
$Z4J$ process and $ZW2J$. The latter has been simulated with \textsc{madgraph} while $Z4J$
has been simulated through a specific\textsc{ alpgen} code for $Z4j$.
Restricting event invariant mass in a $O(100{\rm GeV})$ interval
around $m_{A}$, the total BG cross section is factor 2000
larger than the signal.

{}To increase the S/B ratio we analyse events in a quite analogous way to what was done for H. First we smear jets energy as described above. Then we impose the kinematical cuts\begin{eqnarray}
\Delta R_{JJ,lJ,ll}>0.4,\quad p_{T}^{J}>20{\rm GeV,}\quad  \eta_{J}<2.5,\label{Acuts}\\
|m_{ll}-m_{Z}|<10{\rm GeV,}\quad  p_{T}^{l}>10{\rm GeV,}\quad  \eta_{l}<2.5{\rm .}\nonumber\end{eqnarray}
 The signal events passing these cuts correspond to 3.02(4)fb cross
section while BG cross section is still orders of magnitudes
larger. Thus we impose \emph{reconstruction cuts}. Namely, we
require that the 4 final jets can be divided into 2 pairs reconstructing
 2 vector bosons of the same type. If they are both W, then we require that they
reconstruct an $h.$ If they are both Z, we require that out of the
3 final Z's (the two from jets and the one reconstructed by the leptons)
we should find a  pair reconstructing an $h$. Reconstruction parameters
are the same as in the case of H, Eq. (\ref{deltaV}) and (\ref{deltah}).
After these reconstruction cut signal cross section is $2.2$ fb.
Unfortunately the total BG cross section is still one order
of magnitude larger. However, it's interesting to look closer at the
differential cross sections of BG and signal+BG versus
the event invariant mass, plotted in Fig. \ref{fig:Signal+BGs-and-BG}.
We see that the signal distribution presents a well visible peak above
the BG. The discovery significance can be optimized choosing
a range with largest $S/\sqrt{B}$ ratio. For example, assuming $\mathcal{L}=100 {\rm fb}^{-1}$, in the $595-635$ GeV range we expect 816
events in the SM, and 989 events in $\lambda$SUSY at the benchmark point (\ref{punto}), which amounts to $6.1\sigma$ discovery significance.
\section{Conclusions}
Our conclusion is that $\lambda$SUSY signal (\ref{our-channel})
is indeed observable at the LHC with 100fb$^{-1}$ of integrated
luminosity. If observed, it can provide clean evidence for the heavy
scalar H as well as for the $H\rightarrow hh$ dominant decay chain.
Moreover, we have shown that the CP-odd Higgs boson A
has a clear experimental signature (\ref{Achannel}),
which allows for its discovery at the LHC with 100fb$^{-1}$ of
integrated luminosity. Remarkably, the peaked shape of the signal distribution
should allow BG extraction from data and an easy mass measurement.
Even though the $A\rightarrow Zh$ decay mode is less distinctive
of $\lambda$SUSY than the $H\rightarrow hh$,
its signature seems simpler and cleaner, and it could be the easiest
channel to pursue when looking for $\lambda$SUSY.\begin{acknowledgement}
I would like to thank L. Cavicchia and V.S. Rychkov as co-authors of \cite{CFR}. I would like to thank R. Barbieri, G. Corcella, F. Maltoni, M. Mangano, A. Messina for
useful discussions. I also thank M. Herquet for useful advices on using MadGraph and help
in using MadGraph's cluster.
\end{acknowledgement}





\end{document}